\input{aipcheck.tex}

\def\selectedoptions{}
\ifx\empty\selectedoptions
  \def\selectedoptions{final}
\fi

\documentclass[
   \selectedoptions
  ]
  {aipproc}

\def\selectedlayoutstyle{6x9} 
\layoutstyle\selectedlayoutstyle

\SetInternalRegister\hbadness{8000} % pseudo latin isn't breaking very well :-)

\newcommand\doingARLO[2][]{%
  \ifx\mmref\undefined #1\else #2\fi
}

\begin{document}

\title 
      []
      {Hadronic Decays of Charm}


\author{Kevin Stenson}{
% If you represent a collaboration please use this format
%\author{Author representing the So-and-so collaboration}{
  address={Dept of Physics \& Astronomy, VU Station B351807, Vanderbilt University, Nashville, TN 37235},
  email={stenson@fnal.gov},
  thanks={}
}

%{\centerline Representing the BABAR collaboration}

%
% Typically, these proceedings should be single authored,
% but more authors can be added:
%

%\iftrue
%\author{Arno Mittelbach}{
%  address={Zedernweg 62, 55128 Mainz, Germany},
%  email={arno@mittelbach-online.de},
%}
%
%\author{D. P. Carlisle}{
%  address={Willow House, Souldern},
%  email={david@dcarlisle.demon.co.uk},
%  homepage={http://www.dcarlisle.demon.co.uk},
%  altaddress={When I go to work}{NAG Ltd, Oxford}
%}
%\fi

% \copyrightholder{Acoustical Scociety of America}
\copyrightyear  {2001}

\begin{abstract}
Recent hadronic charm decay results from fixed-target experiments are presented.
New measurements of the $D^0 \rightarrow K^-K^+K^-\pi^+$ branching ratio are shown 
as are recent results from Dalitz plot fits to $D^+ \rightarrow K^-K^+\pi^+\!\!,\;
\pi^+\pi^-\pi^+\!\!,\; K^-\pi^+\pi^+\!\!,\; K^+\pi^-\pi^+$ and $D_s^+ \rightarrow \pi^+\pi^-\pi^+\!\!,\;
 K^+\pi^-\pi^+$.  These fits include measurements of the masses and widths of 
several light resonances as well as strong evidence for the existence of two light scalar 
particles, the $\pi\pi$ resonance $\sigma$ and the $K\pi$ resonance $\kappa$.
\end{abstract}

\date{\today}

\maketitle

\section{Introduction}

Hadronic decays of charm are rich in information about QCD\@.  
For instance, the suppression of $D^0 \rightarrow \pi^+\pi^-$ relative to 
$D^0 \rightarrow K^-K^+$ proved the importance of final state interactions in charm decays.
Also, hadronic decays give rise to the 2.5$\times$ difference between the $D^+$
and $D^0$ lifetimes.  
Hadronic decays can provide information on relative strengths of decay diagrams
(spectator, exchange, annihilation, etc.).  Spectator diagrams are believed to be responsible
for the vast majority of the charm decay rate.  In a spectator diagram, the
charm quark decays while the other quark in the meson is a spectator.  By contrast,
exchange and annihilation diagrams require a connection between the charm quark and the other
quark in the meson and are therefore suppressed.  
Determining the contributions of these diagrams is an interesting open question in charm physics.  

More recently, charm has been used to investigate the light
resonances which are products of charm decays.  Although very high statistics scattering experiments 
have been performed for many years to investigate these resonances, many parameters are still
virtually unknown.  Charm offers a unique way to investigate these resonances by nature of its low
background and well defined initial state (pseudoscalar meson).

Experimentally, hadronic decays can be investigated by comparing branching ratios and by 
analyzing the resonant structure of multibody decays.  The results presented here come from
the Fermilab experiments E791 and FOCUS\@.  E791 (FOCUS) ran in 1991 (\mbox{1996--7}) with a 
500 GeV/$c^2$ $\pi^-$ beam (180 GeV photon beam) on five (four) targets.

\section{Branching ratio measurements}

Calculations of charm \emph{quark} decay rates via the weak interaction have been possible for
many years.  Unfortunately, only charm \emph{hadron} decay rates (which are affected by the
strong force) can be measured by experiments.
The strong force is even more intimately involved when the charm particle decays into hadrons.
Thus, deviations from the na\"\i ve weak prediction for a given decay can provide insight
into the nature of the strong force.

\subsection{$D^0 \!\rightarrow K^-K^+K^-\pi^+$ branching ratio}

\begin{figure}
 \includegraphics[width=.498\textwidth,height=1.67in]{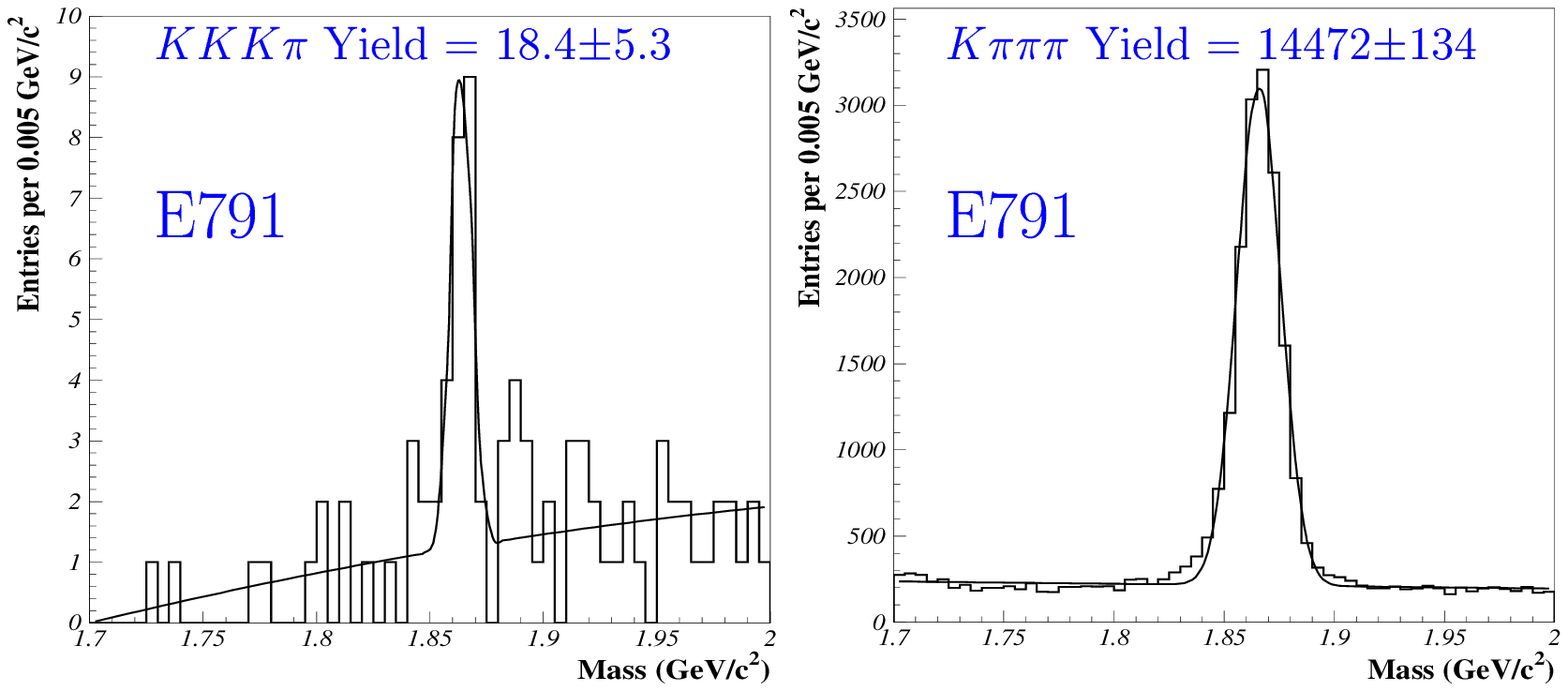}\hspace{0.004\textwidth}
 \includegraphics[width=.249\textwidth,height=1.67in]{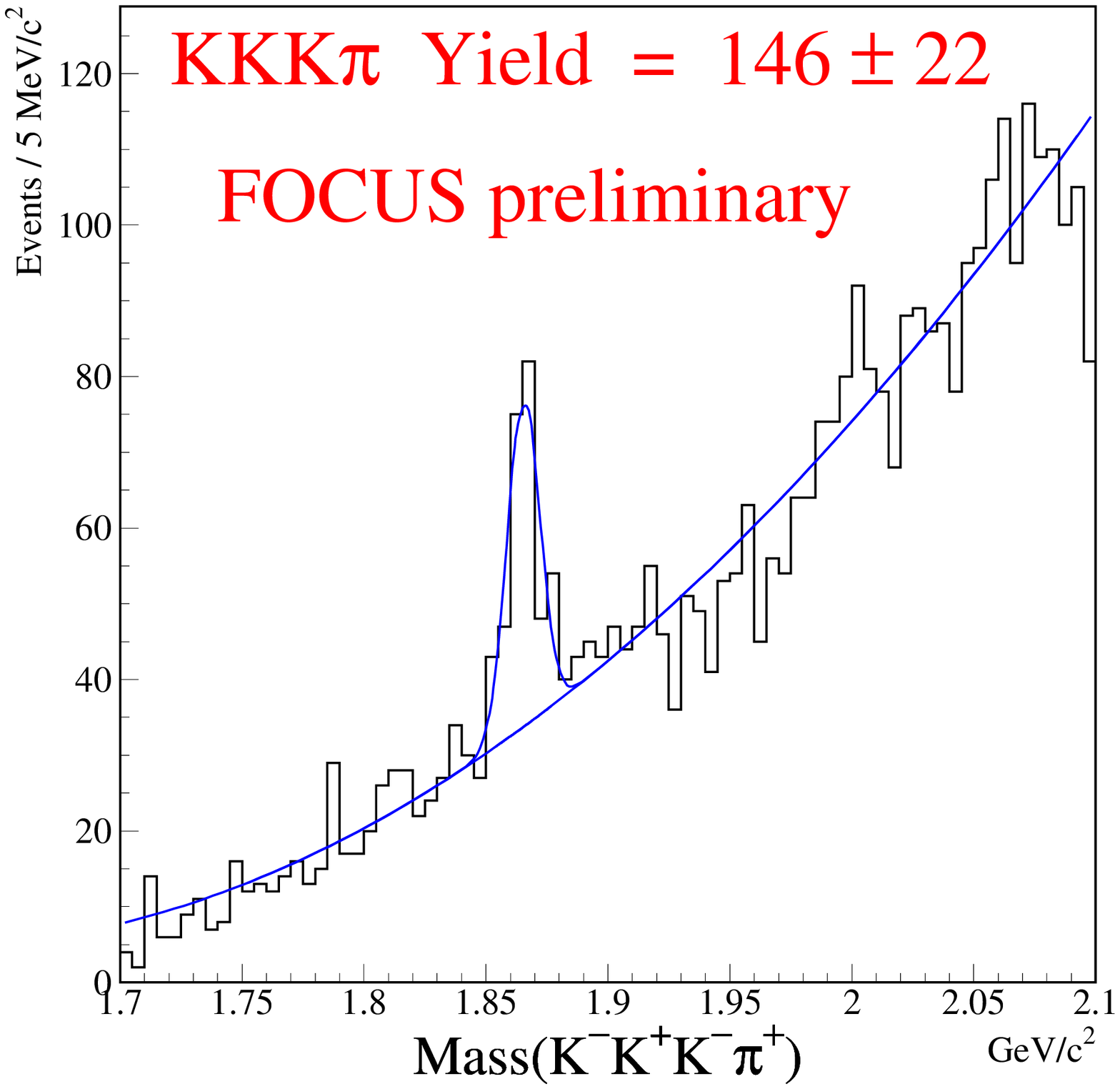}
 \includegraphics[width=.249\textwidth,height=1.67in]{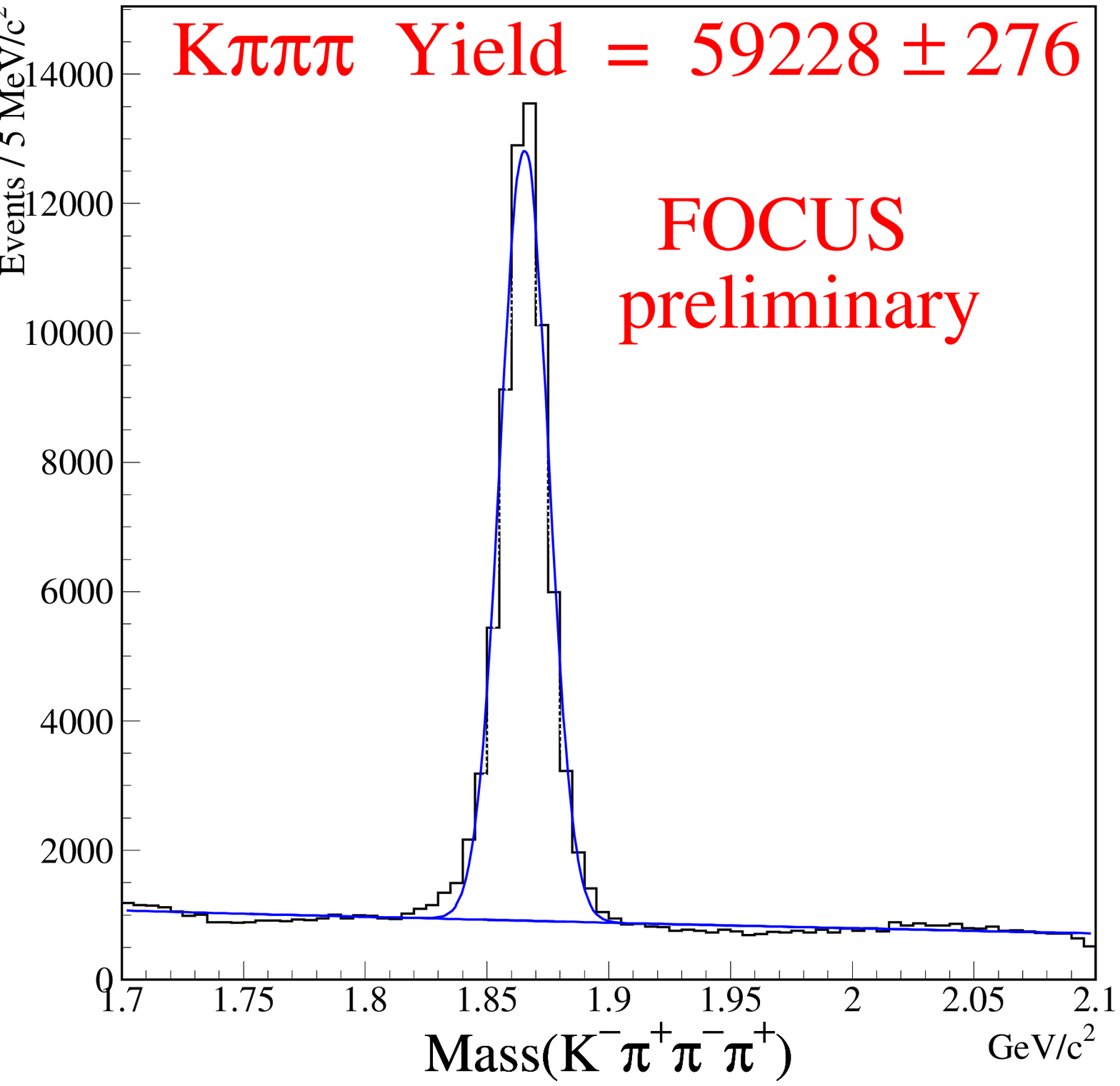}
 \caption{E791 and FOCUS signals from $D^0 \!\rightarrow K^-K^+K^-\pi^+$
 and $D^0 \!\rightarrow K^-\pi^+\pi^-\pi^+$ decays.}
 \label{fig:3kpi}
\end{figure}

The decays $D^0 \!\rightarrow\! K^-K^+K^-\pi^+$ and $D^0 \!\rightarrow\! K^-\pi^+\pi^-\pi^+$ are both
Cabibbo favored decays.  
%That is, Feynman diagrams representing the weak decays can be drawn 
%with W-vertices which only contain $\cos{\theta_C}$ terms.  
A Cabibbo-favored hadronic $D^0(c\bar{u})$ decay produces one $s$ quark from the charm,
one $\bar{u}$ spectator quark, and a $u$ and $\bar{d}$ quark from the virtual W.  If both decay
modes were entirely non-resonant, forming
the $K^-\pi^+\pi^-\pi^+$ final state would require popping $u\bar{u}d\bar{d}$ from the vacuum while
$K^-K^+K^-\pi^+$ would require $u\bar{u}s\bar{s}$ and the branching ratio
between $D^0 \!\rightarrow\! K^-K^+K^-\pi^+$ and $D^0 \!\rightarrow\! K^-\pi^+\pi^-\pi^+$
%BR$\left(\frac{D^0 \rightarrow K^-K^+K^-\pi^+}{D^0 \rightarrow K^-\pi^+\pi^-\pi^+}\right)$
could be used to determine the $s\bar{s}$ suppression relative to $d\bar{d}$.  Multi-body charm decays
generally proceed through resonances, however, which complicates the issue.  Even in this case,
one can note that the $D^0 \!\rightarrow\! K^-\pi^+\pi^-\pi^+$ decay can occur through resonances
using only the four quarks from the decay, 
\textit{e.g.} $\overline{K}^*(892)^0(s\bar{d}) \;\rho(770)^0(u\bar{u})$, while the 
$D^0 \!\rightarrow\! K^-K^+K^-\pi^+$ decay requires either an $s\bar{s}$ pair from the vacuum or 
a final state interaction which couples $\pi\pi$ to $K\overline{K}$.
The recent E791 result contains some of these speculations~\cite{ref:e791_3kpi}.  
Signals for these two modes from E791 and FOCUS are shown in Fig.~\ref{fig:3kpi}.
E791 finds BR$\left(\frac{D^0 \!\rightarrow K^-K^+K^-\pi^+}{D^0 \!\rightarrow K^-\pi^+\pi^-\pi^+}\right)
=0.0054 \pm 0.0016 \pm 0.0008$, significantly higher than the E687 measurement of
$0.0028 \pm 0.0007 \pm 0.0001$~\cite{ref:e791_3kpi,ref:e687_3kpi}.  The preliminary FOCUS result is much closer
to the E687 result at $0.00306 \pm 0.00047$ (statistical error only).  
%Assuming a negligible systematic error on the FOCUS
%result, the world average would be $0.00309 \pm 0.00038$.  
%Translating this number into a 
%result on $s\bar{s}$ suppression, however, is complicated by the resonant substructure of the
%decays which can significantly change the branching fraction of either decay.  A more precise
%analysis using a coherent resonance framework is needed to fully understand the nature of this
%decay.

\section{Dalitz plot analyses}

Multibody decays of charm particles can occur through various strong resonances which can
interfere with each other.  In a three-body decay, a ``Dalitz'' plot can be constructed to 
show the effect of these resonances and their interferences by plotting the 
squared invariant mass of two combinations of the final state particles against each other.
In the absence of interference, how a resonance appears on the Dalitz will depend on its
mass and width (a relativistic Breit-Wigner) 
as well as on its spin (Legendre polynomials).  Interference effects can significantly alter these
shapes.  Fitting a Dalitz plot to a fully coherent sum of resonances allows one to extract
information about how much each resonance contributes and how each resonance interferes 
with other resonances.    By performing a coherent Dalitz
plot analysis one can extract information about weak decays and the effects of the strong
force on the weak decays.

\subsection{$D^+ \!\rightarrow K^-K^+\pi^+$ decays}

\begin{figure}
 \includegraphics[width=\textwidth,height=2.14in]{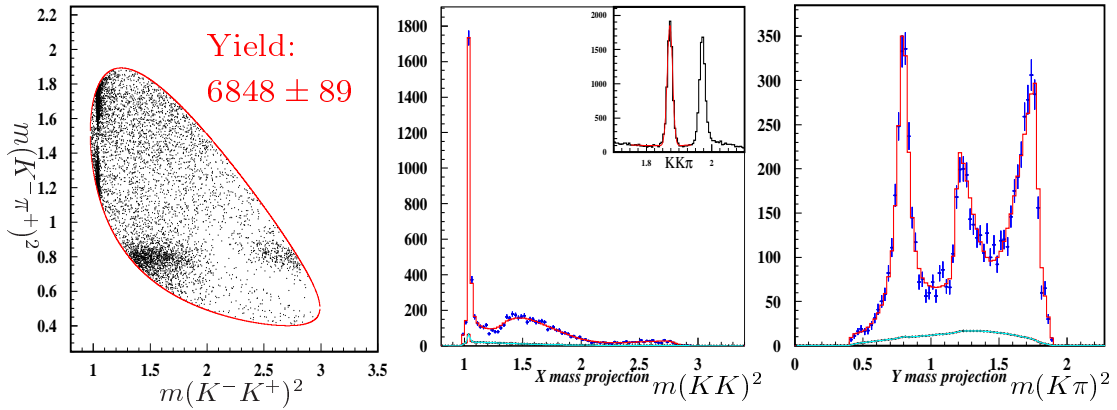}
 \caption{Preliminary $D^+ \!\rightarrow K^-K^+\pi^+$ FOCUS Dalitz plot and projections. 
Projections show the data background, data signal+background, and the fitted result.}
 \label{fig:focus:dpkkpi}
\end{figure}

FOCUS has a high statistics sample of the singly Cabibbo suppressed decay $D^+ \!\rightarrow K^-K^+\pi^+$.
The Dalitz plot as well as projections along each axis for this mode are shown in Fig.~\ref{fig:focus:dpkkpi}.  
In the Dalitz plot and on the $m^2(KK)$ projection, a very clear $\phi(1020)$ can be seen.
The $K^*(892)$ across the Dalitz plot and along the $m^2(K\pi)$ projection is also clear.  
Both contributions show a $\cos{\theta}$ modulation of the amplitude due to the spin-1
nature of the resonances.  Distortions due to interferences are also visible.  
Preliminary fit results from a fully 
coherent analysis of the data are tabulated in Table~\ref{tab:focus:dpkkpi}.  These results
support the presence of significant amounts of $K^*(892)$ and $\phi(1020)$.  The large contribution
from the $K_0^*(1430)$ partly explains the broad enhancement at high $m^2(K\pi)$.  The
existance of many resonances with very different phases indicates significant interferences, as
does the fact that the fit fraction sum is much greater than 100\%.  These interferences explain the 
obvious distortions seen in the Dalitz plot.  Work is currently underway  
to investigate direct \textit{CP} violation by comparing the Dalitz plots of the $D^+$ and $D^-$ decays.  
While for two-body decays a simple branching ratio is sufficient, for multi-body modes a Dalitz plot 
analysis is necessary to extract all of the information on direct \textit{CP} violation.

\begin{table}
 \begin{tabular}{lclcl}
 \hline
 \tablehead{1}{l}{b}{Decay mode} & \ \ \ \ \ & \tablehead{1}{l}{b}{Fraction (\%)} & \ & \tablehead{1}{l}{b}{Phase ($^\circ$)} \\ \hline
 $K^*(892) K^+$ & & $22.0 \pm 1.1$ & & $0$ (fixed) \\
 $a_0(980) \pi^+$ & & $27.8 \pm 4.8$ & & $146 \pm 5$ \\
 $\phi(1020) \pi^+$ & & $27.8 \pm 0.9$ & & $244 \pm 6$ \\
 $f_2(1270) \pi^+$ & & $0.7 \pm 0.2$ & & $12 \pm 7$ \\
 $f_0(1370) \pi^+$ & & $5.9 \pm 1.2$ & & $60 \pm 6$ \\
 $K^*(1410) K^+$ & & $8.8 \pm 1.9$ & & $135 \pm 6$ \\
 $K_0^*(1430) K^+$ & & $69.3 \pm 6.3$ & & $63 \pm 4$ \\
 $\phi(1680) \pi^+$ & & $1.5 \pm 0.5$ & & $-70 \pm 9$ \\ %\hline
 sum & & 163.8 & \\ \hline
 \end{tabular}
 \caption{Preliminary FOCUS Dalitz plot fit results for the decay $D^+ \!\rightarrow K^-K^+\pi^+$. 
  (Statistical errors only.)}
 \label{tab:focus:dpkkpi}
\end{table}

%\subsection{$D^+, D_s^+ \rightarrow \pi^+\pi^-\pi^+$ decays}
%
%The E791 mass and Dalitz plots for the $\pi^+\pi^-\pi^+$ final state are shown in Fig.~\ref{fig:e791:3pi_dal}.
%Using the signals shown in Fig.~\ref{fig:e791:3pi_dal}, E791 measures $D^+\rightarrow \pi^-\pi^+\pi^+$ and
%$D_s^+\rightarrow \pi^-\pi^+\pi^+$ branching ratios shown in Table~\ref{tab:e791:3pi_br}~\cite{ref:e791_ds3pi}.
%
%\begin{table}
% \begin{tabular}{lll}
% \hline
% \tablehead{1}{c}{b}{Branching Ratio} & \tablehead{1}{c}{b}{E791} & \tablehead{1}{c}{b}{PDG Fit~\cite{ref:pdg2k}} \\ \hline
%$\Gamma \left(\frac{D^+\rightarrow \pi^-\pi^+\pi^+}{D^+\rightarrow K^-\pi^+\pi^+}\right) $ & 
%$\left(3.11 \pm 0.18 ^{\,+\,0.16}_{\,-\,0.26}\right)\%$ & $\left(4.06 \pm 0.34\right)\%$ \\
%$\Gamma \left(\frac{D_{s}^+\rightarrow \pi^-\pi^+\pi^+}{D_{s}^+\rightarrow \phi\pi^+}\right) $ & 
%$\left(24.5 \pm 2.8 ^{\,+\,1.9}_{\,-\,1.2}\right)\%$ & $\left(28 \pm 6\right)\%$ \\ \hline
% \end{tabular}
% \caption{E791 branching ratio results for the decays $D^+ \rightarrow \pi^+\pi^-\pi^+$ and 
% $D_s^+ \rightarrow \pi^+\pi^-\pi^+$.}
% \label{tab:e791:3pi_br}
%\end{table}

\subsection{$D_s^+ \rightarrow \pi^+\pi^-\pi^+$ decays}

\begin{figure}
 \hspace{0.05\textwidth}\includegraphics[width=0.35\textwidth,height=0.55in]{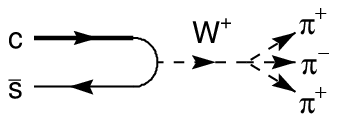}\hspace{0.2\textwidth}
 \includegraphics[width=0.35\textwidth,height=0.55in]{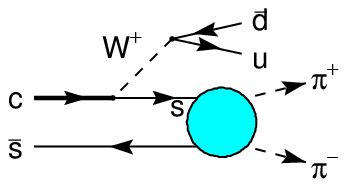}\hspace{0.05\textwidth}
 \caption{Diagram of annihilation (left) and resonance (right) contributions to $D^+_s \!\rightarrow \pi^+\pi^-\pi^+$.}
 \label{fig:feyn_3pi}
\end{figure}

Although $D_s^+ \!\!\rightarrow\! \pi^+\pi^-\pi^+$ is Cabibbo favored, it can only occur via a
spectator diagram if it uses a resonance which couples to both $K\overline{K}$ and $\pi\pi$ or
via final state interactions.  It can also proceed via an annihilation diagram.
These possibilities are sketched in Fig.~\ref{fig:feyn_3pi}.  Since $\rho(770)$ does
not couple to $K\overline{K}$, any $\rho(770)$ would indicate an
annihilation diagram contribution or final state interactions.  A resonance known to couple 
to $K\overline{K}$ and $\pi\pi$ is the $f_0(980)$.  This mysterious state 
has been proposed as a four-quark state and $K\overline{K}$ molecule among other
things.  The presence of the $a_0(980)$ further complicates the understanding
of this unique resonance.  The $f_0(980)$ mass is below the $K\overline{K}$ threshold but
is broad enough to have a significant branching fraction to $K\overline{K}$ even with limited phase space.

\begin{figure}
 \includegraphics[width=\textwidth,height=1.83in]{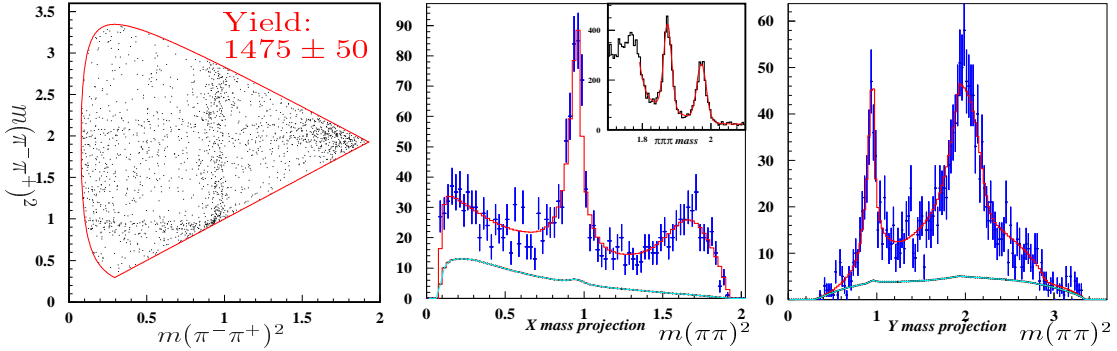}
 \caption{Preliminary  $D_s^+ \!\rightarrow \pi^+\pi^-\pi^+$ FOCUS Dalitz plot and projections.
Projections show the data background, data signal+background, and the fitted result.}
 \label{fig:focus:dspipipi}
\end{figure}

The FOCUS Dalitz plot and projections for $D_s^+ \rightarrow \pi^+\pi^-\pi^+$ are shown in 
Fig.~\ref{fig:focus:dspipipi}.  Clear $f_0(980)$ bands are visible in the 
Dalitz plot and projections.  A concentration at $m^2(\pi^+\pi^-) \sim 2~\mbox{GeV/}c^2$ is also visible.  
The results of a preliminary fit to this distribution as well
as E791 results~\cite{ref:e791_ds3pi} are shown in Table~\ref{tab:dspipipi}.
Both results clearly show $f_0(980)$ dominance and no significant
$\rho(770)$.  Thus, this decay proceeds almost entirely through resonance modes
with no evidence of an annihilation diagram contribution.

%\begin{figure}
% \includegraphics[width=0.99\textwidth,height=2.0in]{e791_3pidal.eps}
% \caption{E791 mass and Dalitz plots of the decays $D^+ \rightarrow \pi^+\pi^-\pi^+$ and 
% $D_s^+ \rightarrow \pi^+\pi^-\pi^+$ }
% \label{fig:e791:3pi_dal}
%\end{figure}

\begin{table}
 \begin{tabular}{lcllcll}
 \hline
  & & \tablehead{2}{c}{b}{E791} & & \tablehead{2}{c}{b}{FOCUS\ \ (preliminary)} \\
 \tablehead{1}{l}{b}{Decay mode} & \ \ \ \ \ \ \ \ & 
 \tablehead{1}{l}{b}{Fraction (\%)} & \tablehead{1}{l}{b}{Phase ($^\circ$)} &  \ \ \ \ \ \ & 
 \tablehead{1}{l}{b}{Fraction (\%)} & \tablehead{1}{l}{b}{Phase ($^\circ$)} \\ \hline
 $f_0(980) \pi^+$ & & $56.5 \pm 4.3 \pm 4.7$ & $0$ (fixed) & & $94.4 \pm 2.5$ & 0 (fixed) \\ 
 non-resonant                   & & $0.5 \pm 1.4 \pm 1.7$ & $181 \pm 94 \pm 51$ & & $25.5 \pm 4.4$ & $246 \pm 4$ \\
 $\rho(770) \pi^+\!$ & & $5.8 \pm 2.3 \pm 3.7 $ & $109 \pm 24 \pm 5$ & & & \\
 $f_2(1270) \pi^+$ & & $19.7 \pm 3.3 \pm 0.6$ & $133 \pm 13 \pm 28 $ & & $9.8 \pm 1.2$ & $140 \pm 6$ \\
 $\rho(1450) \pi^+$ & & $4.4 \pm 2.1 \pm 0.2$ & $162 \pm 26 \pm 17$ & & $4.1 \pm 0.7$ & $188 \pm 14$ \\
 $S_0(1475) \pi^+$ & & & & & $17.4 \pm 2.2$ & $250 \pm 4$ \\
 $f_0(1370) \pi^+\!$ & & $32.4 \pm 7.7 \pm 1.9$ & $198 \pm 19 \pm 27$ & & & \\ %\hline
 sum & & 119.3 & & & 151.2 & \\ \hline
 \end{tabular}
 \caption{E791~\cite{ref:e791_ds3pi} and preliminary FOCUS Dalitz plot fit results for the decay 
$D_s^+ \rightarrow \pi^+\pi^-\pi^+$.  Errors on FOCUS results are statistical only.}
 \label{tab:dspipipi}
\end{table}

%In the course of fitting these Dalitz plots, it was noticed that
%many resonances are either poorly measured or have significant discrepancies between different
%experiments.  The resonance parameters often come from scattering experiments which suffer from
%significant backgrounds and poorly defined initial/final states.  In contrast, the charm system is 
%a clean environment with well known inital and final states which can be used to study the light 
%resonances.  
Using data, E791 and FOCUS have made measurements of
the poorly measured scalars contributing to the $D_s^+ \rightarrow \pi^+\pi^-\pi^+$ decay.
Although the PDG~\cite{ref:pdg2k} lists $f_0(980)$ mass and width measurements from 1973 to the 
present, with 13 measurements in 1999 alone, there is still no consensus on either value.  The I=0, J=0
states betweeen 1200 and 1500 MeV/$c^2$ are even more murky.

E791 finds a mass and width of the $f_0(1370)$ of $1434 \pm 18 \pm 9$~MeV/$c^2$ and 
$172 \pm 32 \pm 6$ MeV/$c^2$, respectively.  FOCUS uses $S_0(1475)$ for a 
scalar around 1475~MeV/$c^2$, as seen in an E687 analysis.  The preliminary mass and width are 
found to be $1473 \pm 8$ MeV/$c^2$ and $112 \pm 17$ MeV/$c^2$, respectively; quite comparable to 
the E791 result for $f_0(1370)$. 

E791 finds slightly better fits using the WA92 coupled channel Breit-Wigner formalism to describe
the $f_0(980)$.  A standard relativistic Breit-Wigner is proportional to 
1/$(m^2 - m_r^2 + im_r\Gamma_{r}(m^2))$ for a resonance $r$ with a mass and width of $m_r$ and
$\Gamma_r$, respectively and at a two-body mass-squard of $m^2$.  In the $f_0(980)$ WA92 coupled channel 
formula, $\Gamma_r$ is replaced with $\Gamma^\pi_r + \Gamma^K_r$ 
where $\Gamma^\pi_r(m^2) = g_\pi \sqrt{m^2_{\pi\pi}/4 - m_\pi^2}$ and 
$\Gamma^K_r(m^2) = g_K \left( \sqrt{m^2_{\pi\pi}/4 - m_{K^+}^2} + \sqrt{m^2_{\pi\pi}/4 - m_{K^0}^2}\right)$.
This framework yields $M_{f_0(980)} = 977 \pm 3 \pm 2$  MeV/$c^2$, 
$g_\pi = 0.09 \pm 0.01 \pm 0.01$, and $g_K = 0.02 \pm 0.04 \pm 0.03$ from the E791 data.  
Fitting with a
standard Breit-Wigner results in minor changes: $M_{f_0(980)} = 975 \pm 3 \pm 2$~MeV/$c^2$ and
$\Gamma_{f_0(980)} = 44 \pm 2 \pm 2$  MeV/$c^2$.
FOCUS finds the K-matrix framework works quite well in dealing with the coupled channel nature of
the $f_0(980)$~\cite{ref:sandra,ref:kmatrix}.  In this framework, transformed variables are used:
$m_0^2 = m_r^2 + (\gamma_{KK}/\gamma_{\pi\pi})^2 (|\rho_{KK}(m_r)|/\rho_{\pi\pi}(m_r)) m_r \Gamma_r$
and $\Gamma_0 = m_r \Gamma_r / (m_0 \rho_{\pi\pi}(m_r) \gamma^2_{\pi\pi})$ where $\rho_{\pi\pi}$ and
$\rho_{KK}$ are phase space terms and $\gamma_{\pi\pi}$ and $\gamma_{KK}$ are coupling constants 
normalized to $\gamma^2_{\pi\pi} + \gamma^2_{KK} = 1$.  In this framework, the preliminary FOCUS fits
return $M_{f_0(980)} = 963 \pm 6$ MeV/$c^2$, $\Gamma_{f_0(980)} = 297 \pm 92$ MeV/$c^2$, and
$\gamma^2_{KK}/\gamma^2_{\pi\pi} = 2.09 \pm 0.53$ which translates to
$M_{f_0(980)} = 982 \pm 30$ MeV/$c^2$ and $\Gamma_{f_0(980)} = 89 \pm 32$ MeV/$c^2$ for a 
standard Breit Wigner (errors are statistical only).

\subsection{$D^+ \rightarrow \pi^+\pi^-\pi^+$}

E791 has published results for a coherent Dalitz plot analysis of 
$D^+ \rightarrow \pi^+\pi^-\pi^+$~\cite{ref:e791_dp3pi}.  In their initial fit to the Dalitz plot
using all known resonances, the fit quality was very poor with a confidence level of $10^{-5}$.  
By including a low mass scalar particle ($\sigma$) the fit was significantly improved and
yielded a confidence level of 75\%, providing strong evidence for the elusive light scalar.
The results from both fits are tabulated in Table~\ref{tab:e791:dp3pi}.
From the fit, the $\sigma$ parameters were determined to be $M_{\sigma} = 478^{\,+\,24}_{\,-\,23} \pm 17$~MeV/$c^2$
and $\Gamma_\sigma = 324^{\,+\,42}_{\,-\,40} \pm 21$~MeV/$c^2$.  Many checks were made to validate the existence
of the $\sigma$ in this decay.  These checks included fitting with a vector and tensor state, and fitting 
with no phase variation.  The fit with a phase-varying scalar particle was clearly preferred.

\begin{table}[htb]
 \begin{tabular}{lllll}
 \hline
  & \tablehead{2}{c}{b}{Without $\sigma$: CL = 0.001\%} & 
    \tablehead{2}{c}{b}{With $\sigma$: CL = 75\%} \\
 \tablehead{1}{l}{b}{Decay mode} & \tablehead{1}{l}{b}{Fraction (\%)} & \tablehead{1}{l}{b}{Phase ($^\circ$)} & 
 \tablehead{1}{l}{b}{Fraction (\%)} & \tablehead{1}{l}{b}{Phase ($^\circ$)} \\ \hline
$\rho(770) \pi^+\!$ & $20.8 \pm 2.4 $ & $0$ (fixed) & $33.6 \pm 3.2 \pm 2.2 $ & $0$ (fixed) \\
non-resonant & $38.6 \pm 9.7$ & $150 \!\pm 12$ & $7.8 \pm 6.0 \pm 2.7$ & $57 \pm 20 \pm 6$ \\
$f_0(980) \pi^+\!$ & $7.4 \pm 1.4$ & $152 \pm 16$ & $6.2 \pm 1.3 \pm 0.4$ & $165 \pm 11 \pm 3$ \\
$f_2(1270) \pi^+\!$ & $6.3 \pm 1.9$ & $103 \pm 16$ & $19.4 \pm 2.5 \pm 0.4$ & $57 \pm 8 \pm 3 $ \\
$f_0(1370) \pi^+\!$ & $10.7 \pm 3.1$ & $143 \pm 10$ & $2.3 \pm 1.5 \pm 0.8$ & $105 \pm 18 \pm 1$ \\
$\rho(1450) \pi^+\!$ & $22.6 \pm 3.7$ & $46 \pm 15$ & $0.7 \pm 0.7 \pm 0.3$ & $319 \pm 39 \pm 11$ \\
$\sigma\, \pi^+\!$ &  &  & $46.3 \pm 9.0 \pm 2.1$ & $206 \pm 8 \pm 5$ \\ %\hline
sum & 106.4 & & 116.3 & \\ \hline
 \end{tabular}
 \caption{E791 Dalitz plot fit results for the decay 
$D^+ \rightarrow \pi^+\pi^-\pi^+$~\cite{ref:e791_dp3pi}. Parameter errors from ``without $\sigma$'' fits are statistical only.}
 \label{tab:e791:dp3pi}
\end{table}

\subsection{$D^+ \rightarrow K^- \pi^+ \pi^+$}

The Cabibbo favored decay $D^+ \rightarrow K^- \pi^+ \pi^+$ provides a very high statistics mode in which
to study charm decays.  
Previous analyses of this decay~\cite{ref:e691_k2pi,ref:e687_k2pi} 
have identified two mysteries in this decay.  The first mystery is why there is a
dominant non-resonant contribution; unique in charm decays.  The second mystery is why a good fit to this Dalitz
plot seems to be impossible to achieve.
%This mode contains a major mystery in charm decay physics, that is, the dominant
%contribution of the decay through a non-resonant process; unique in charm decays.  
The E791 data sample of 15,090 events (94\% signal) can be used to shed light on these mysteries.
Fitting the data using all known resonances results in a large non-resonant contribution and a very
poor fit (confidence level of 10$^{-11}$), similar to past attempts.  Given the evidence for the 
$\sigma$ in the $D^+ \rightarrow \pi^+\pi^-\pi^+$ mode, an additional scalar ($\kappa$) was added to 
the $D^+ \rightarrow K^- \pi^+ \pi^+$ fit.  This provides a dramatic reduction in the non-resonant
contribution (from 91\% to 13\%) and a much improved fit (confidence level of 95\%).
The preliminary results of both fits are shown in Table~\ref{tab:e791:dpkpipi}.  The preliminary $\kappa$ 
parameters are found to be $M_\kappa = 797 \pm 19 \pm 42$~MeV/$c^2$ and 
$\Gamma_\kappa = 410 \pm 43 \pm 85$~MeV/$c^2$.  Attempts to explain the data in other ways (\textit{e.g.} 
including a vector state, tensor state, non-phase-varying state, structured non-resonant contribution) have been 
inadequate.  Preliminary meausrements of the $K^*_0(1430)$ parameters ($M_{K^*_0(1430)} = 1459 \pm 7 \pm 6$~MeV/$c^2$ 
and $\Gamma_{K^*_0(1430)} = 175 \pm 12 \pm 12$~MeV/$c^2$) have also been extracted in this 
analysis~\cite{ref:carla}.

\begin{table}[hb]
 \begin{tabular}{lllll}
 \hline
  & \tablehead{2}{c}{b}{Without $\kappa$: CL = 10$^{-11}$} & 
    \tablehead{2}{c}{b}{With $\kappa$: CL = 95\%} \\
 \tablehead{1}{l}{b}{Decay mode} & \tablehead{1}{l}{b}{Fraction (\%)} & \tablehead{1}{l}{b}{Phase ($^\circ$)} & 
 \tablehead{1}{l}{b}{Fraction (\%)} & \tablehead{1}{l}{b}{Phase ($^\circ$)} \\ \hline
NR                & $90.9 \pm 2.6$ & 0 (fixed) & $13.0 \pm 5.8 \pm 2.6$ & $349 \pm 14 \pm 8$ \\
$\overline{K}^*(892) \pi^+$ & $13.8 \pm 0.5$ & $54 \pm 2$ & $12.3 \pm 1.0 \pm 0.9$ & 0 (fixed) \\
$\overline{K}^*_0(1430) \pi^+$ & $30.6 \pm 1.6$ & $54 \pm 2$ & $12.5 \pm 1.4 \pm 0.4$ & $48 \pm 7 \pm 10$ \\
$\overline{K}^*_2(1430) \pi^+$ & $0.4 \pm 0.1$ & $33 \pm 8$ & $0.5 \pm 0.1 \pm 0.2$ & $306 \pm 8 \pm 6$ \\
$\overline{K}^*(1680) \pi^+$ & $3.2 \pm 0.3$ & $66 \pm 3$ & $2.5 \pm 0.7 \pm 0.2$ & $28 \pm 13 \pm 15$ \\
$\kappa\, \pi^+$ & & & $47.8 \pm 12.1 \pm 3.7$ & $187 \pm 8 \pm 17$ \\ %\hline
sum & 138.9 & & 88.6 & \\ \hline
 \end{tabular}
 \caption{Preliminary E791 Dalitz plot fit results for the decay 
$D^+ \rightarrow K^-\pi^+\pi^+$. Parameter errors from ``without $\kappa$'' fits are statistical only.}
 \label{tab:e791:dpkpipi}
\end{table}

\subsection{$D^+, D_s^+ \rightarrow K^+ \pi^- \pi^+$}

\begin{figure}
\begin{minipage}{\textwidth}
 \includegraphics[width=\textwidth,height=1.725in]{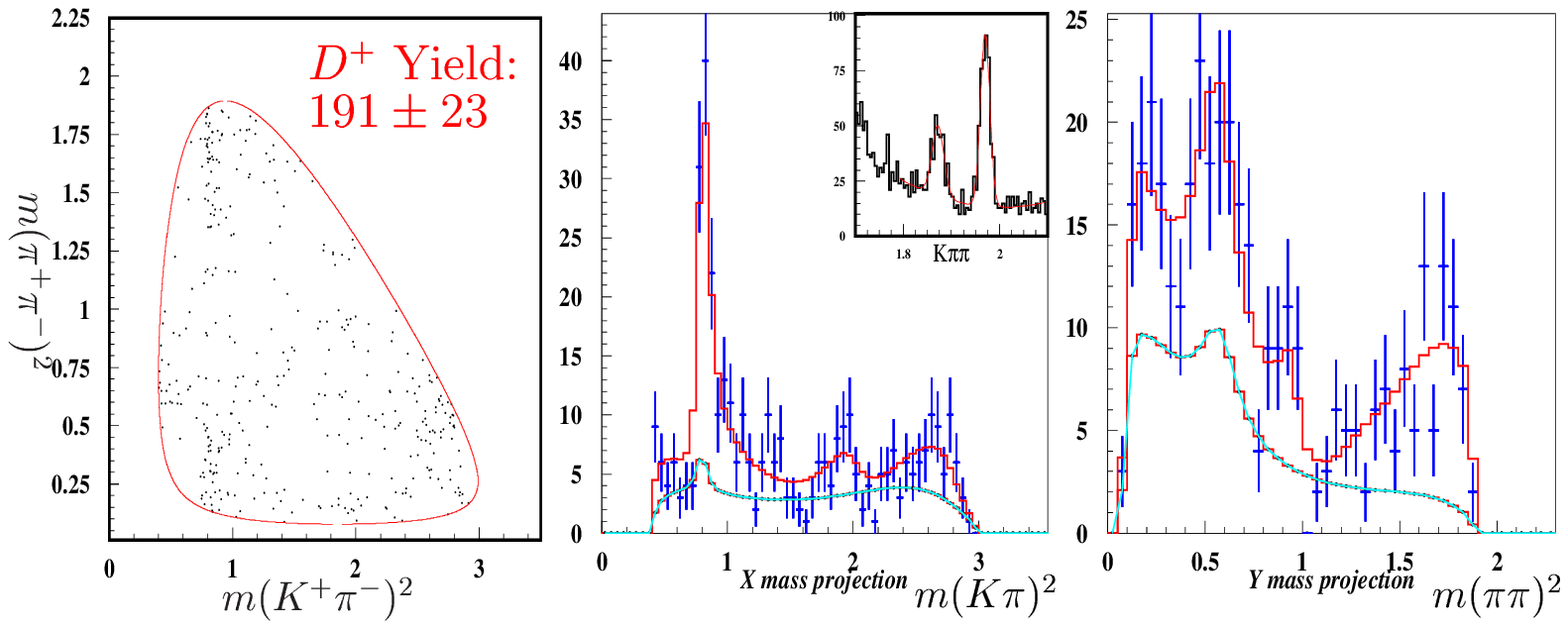}\\
 \includegraphics[width=\textwidth,height=1.725in]{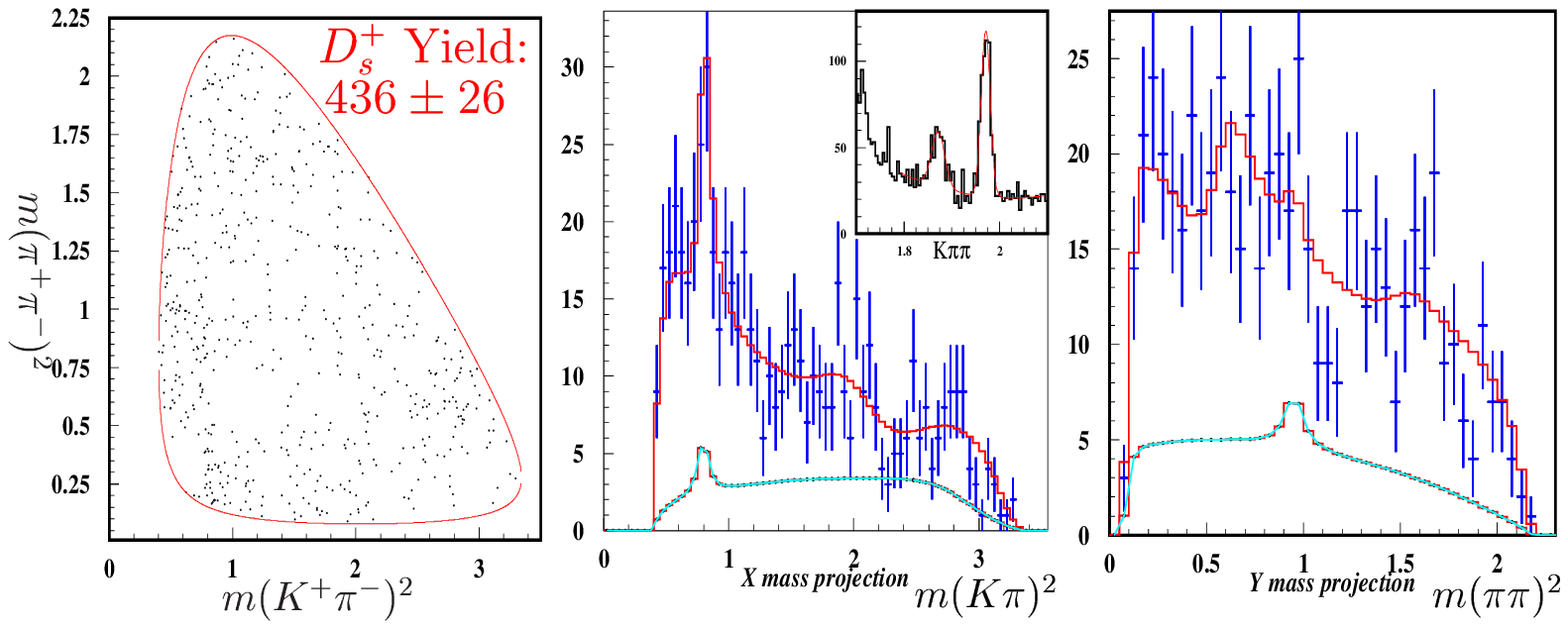}
\end{minipage}
 \caption{FOCUS Dalitz plots and projections for $D^+ \rightarrow K^+ \pi^- \pi^+$ and 
  $D_s^+ \!\rightarrow K^+ \pi^- \pi^+$.
Projections show the data background, data signal+background, and the fitted result.}
 \label{fig:focus:kpipi}
\end{figure}

FOCUS has obtained preliminary results from Dalitz plot fits to the doubly Cabibbo suppressed decay 
$D^+ \rightarrow K^+ \pi^- \pi^+$ and the singly Cabibbo suppressed decay 
$D_s^+ \rightarrow K^+ \pi^- \pi^+$.  The Dalitz plots and projections are shown
in Fig.~\ref{fig:focus:kpipi} and the fit results in Table~\ref{tab:focus:kpipi}.
This is the first fit to the $D_s^+ \rightarrow K^+ \pi^- \pi^+$ Dalitz plot.
%measurement is the first one made for this mode while the $D^+$ is of much higher
%precision than the previous analysis.
%The $D^+$ measurement is much higher precision than the previous analysis while the $D_s^+$
%measurement is the first one made for this mode.  
Both fit results indicate a rich resonance
structure, dominated by $\rho(770)$.

\begin{table}
 \begin{tabular}{lcllcll}
 \hline
  & \ \ \ \ \ \ & \tablehead{2}{c}{b}{$D^+ \rightarrow K^+ \pi^- \pi^+$} & \ \ &
    \tablehead{2}{c}{b}{$D_s^+ \rightarrow K^+ \pi^- \pi^+$} \\
 \tablehead{1}{l}{b}{Decay mode} & & \tablehead{1}{l}{b}{Fraction (\%)$\!\!\!$} & \tablehead{1}{l}{b}{Phase ($^\circ$)} & &
 \tablehead{1}{l}{b}{Fraction (\%)$\!\!\!$} & \tablehead{1}{l}{b}{Phase ($^\circ$)}\ \\ \hline
$\rho(770) K^+$ & & $51 \pm 10$ & 0 (fixed) & & $40 \pm 4$ & 0 (fixed) \\
non-resonant   & & $9 \pm 5$ & $-6 \pm 16$ & & $18 \pm 4$ & $34 \pm 7$ \\
$K^*(892) \pi^+$ & & $43 \pm 7$ & $208 \pm 16$ & & $22 \pm 3$ & $163 \pm 7$ \\
$f_0(980) K^+$ & & $9 \pm 5$ & $73 \pm 31$ & & & \\
$f_2(1270) K^+$ & & & & & $2 \pm 1$ & $33 \pm 21$ \\
$K^*(1410) \pi^+$ & & $12 \pm 8$ & $133 \pm 23$ & & $14 \pm 5$ & $-10 \pm 7$ \\
$K^*_0(1430) \pi^+$ & & & & & $14 \pm 6$ & $68 \pm 7$ \\
$K^*_2(1430) \pi^+$ & & $6 \pm 3$ & $48 \pm 27$ & & & \\
$\rho(1450) K^+$ & & $10 \pm 5$ & $247 \pm 15$ & & $8 \pm 2$ & $219 \pm 14$ \\
$K^*(1680) \pi^+$ & & $22 \pm 10$ & $2 \pm 20$ & & & \\ %\hline
sum & & 162 & & & 118 & \\ \hline
 \end{tabular}
 \caption{Preliminary FOCUS Dalitz plot fit results for
$D^+ \!\!\rightarrow\! K^+ \pi^- \pi^+$ and $D_s^+ \!\!\rightarrow\! K^+ \pi^- \pi^+$. Statistical errors only.}
 \label{tab:focus:kpipi}
\end{table}

\section{Conclusion}

Hadronic decays of charm provides an environment to study many aspects of high energy physics including
measuring the contributions of various Feynman diagrams, studying the effect of final state interactions, searching
for \textit{CP} violation, and measuring the mass and width of light resonances.  The relatively large decay rate of
$D_s^+ \!\rightarrow \pi^+\pi^-\pi^+$ decay has been found to be due to resonances which couple simultaneously to 
$K\overline{K}$ and $\pi\pi$ rather than annihilation diagrams.  Strong evidence for, and precise measurements of, 
two particles which have existed on the fringe for many years ($\sigma$ and $\kappa$) are presented.
Using the clean charm environment to measure light resonance parameters is a new and interesting use of charm
hadronic decays.
Table~\ref{tab:resval} summarizes all of the measured light resonance values described in these proceedings.  
The future holds
the prospect of even more precise measurements of light resonances, a search for direct \textit{CP} violation from
$D^+ \!\rightarrow K^- K^+ \pi^+$ decays, and much more information from many decay modes.

\begin{table}
 \begin{tabular}{lllll}
 \hline
  & \tablehead{2}{c}{b}{E791} & \tablehead{2}{c}{b}{FOCUS} \\
\tablehead{1}{l}{b}{Resonance} & \tablehead{1}{l}{b}{Mass (MeV/$c^2$)} & \tablehead{1}{l}{b}{$\Gamma$ (MeV/$c^2$)} & 
 \tablehead{1}{l}{b}{Mass (MeV/$c^2$)} & \tablehead{1}{l}{b}{$\Gamma$ (MeV/$c^2$)} \\ \hline
$\sigma$ & $478^{\,+\,24}_{\,-\,23} \pm 17$ & $324^{\,+\,42}_{\,-\,40}\pm 21$ \\
$\kappa$ & $797 \pm 19 \pm 42$ & $410 \pm 43 \pm 85$ \\
$f_0(980)$ & $975 \pm 3 \pm 2$ & $44 \pm 2 \pm 2$ & $982 \pm 30 $ & $89 \pm 32$  \\
$f_0(1370)/S_0(1475)$ & $1434 \pm 18 \pm 9$ & $172 \pm 32 \pm 6$ & $1473 \pm 8$ & $112 \pm 17$ \\
$K^*_0(1430)$ & $1459 \pm 7 \pm 6$ & $175 \pm 12 \pm 12$ \\  \hline
 \end{tabular}
 \caption{Fitted masses and widths for various scalars involved in charm decays.  
E791 $\kappa$ and $K^*_0(1430)$ results are preliminary.  The FOCUS results are preliminary and errors shown are
only statistical.
}
 \label{tab:resval}
\end{table}

%example of a table
%\begin{table}
%\begin{tabular}{lrrrr}
%\hline
%  & \tablehead{1}{r}{b}{Single\\outlet}
%  & \tablehead{1}{r}{b}{Small\tablenote{2-9 retail outlets}\\multiple}
%  & \tablehead{1}{r}{b}{Large\\multiple}
%  & \tablehead{1}{r}{b}{Total}   \\
%\hline
%1982 & 98 & 129 & 620    & 847\\
%1987 & 138 & 176 & 1000  & 1314\\
%1991 & 173 & 248 & 1230  & 1651\\
%1998\tablenote{predicted} & 200 & 300 & 1500  & 2000\\
%\hline
%\end{tabular}
%%\source{Central Statistical Office, UK}
%\caption{Average turnover per shop: by type
%  of retail organisation}
%\label{tab:a}
%\end{table}

% choose bibtex style depending on layout style and options used in
% sample:

\doingARLO[\bibliographystyle{aipproc}]
          {\ifthenelse{\equal{\AIPcitestyleselect}{num}}
             {\bibliographystyle{arlonum}}
             {\bibliographystyle{arlobib}}
          }
\bibliography{writeup}

\end{document}